\documentstyle[11pt,aaspp4]{article}
\def\gsim{\;\rlap{\lower 2.5pt
 \hbox{$\sim$}}\raise 1.5pt\hbox{$>$}\;}
\def\lsim{\;\rlap{\lower 2.5pt
   \hbox{$\sim$}}\raise 1.5pt\hbox{$<$}\;}
\newcommand\beq{\begin{equation}}
\newcommand\eeq{\end{equation}}
\def\lya{Ly$\alpha$~}
\newcommand{\Ref}{\hangindent=20pt \hangafter=1 \noindent}
\newcommand{\StartRef}{\hyphenpenalty=10000 \raggedright
\parskip=0pt \parindent=0pt }

\begin{document}

\title{Constraints from the Hubble Deep Field on High Redshift Quasar
Models}

\author{Zolt\'{a}n Haiman\altaffilmark{1}, Piero Madau\altaffilmark{2}, 
and Abraham Loeb\altaffilmark{1}
\altaffiltext{1}{Astronomy Department, Harvard University, 
60 Garden Street, Cambridge, MA 02138, USA}
\altaffiltext{2}{Space Telescope Science Institute, 3700 San Martin Drive,
Baltimore, MD 21218, USA}}

\begin{abstract}
High resolution, deep imaging surveys are instrumental in setting
constraints on semi--analytical structure formation models in Cold
Dark Matter (CDM) cosmologies.  We show here that the lack of
unresolved B--band ``dropouts'' with $V>25$ mag in the Hubble Deep
Field (HDF) appears to be inconsistent with the expected number of
quasars if massive black holes form with a constant universal
efficiency in all CDM halos.  To reconcile the models with the data, a
mechanism is needed that suppresses the formation of quasars in halos
with circular velocities $v_{\rm circ} \lsim 50-75~{\rm km~s^{-1}}$.
This feedback naturally arises due to the photoionization heating of
the gas by the UV background.  We consider several alternative effects
that would help reduce the quasar number counts, and find that these
can not alone account for the observed lack of detections.  If
reddening by dust can be neglected at early epochs, consistency with
the optical data also requires that the luminous extent of dwarf
galaxies at high redshifts be larger than a few percent of the virial
radii of their dark matter halos, in order not to overpredict the
number of point--like B--band dropouts. Future deep observations in
the $J$ and $H$ bands with NICMOS might reveal several $z\gsim 5$
objects per field or provide even stronger constraints on the models
than existing $B, V$, and $I$ data.
\end{abstract}


\centerline{submitted to the {\it The Astrophysical Journal}, May 1998}

\newpage
\section{Introduction}

One of the primary goals of models for structure formation in the
Universe is to explore the formation process of luminous objects such
as quasars or galaxies, and to explain the shape and evolution of
their observed luminosity functions (LFs).  Recent theoretical work
has focused on the formation of galaxies within the framework of
hierarchical structure formation in CDM cosmologies, employing either
three--dimensional numerical simulations (Gnedin \& Ostriker 1997;
Zheng~et~al.~1997), or semi--analytic modeling (Kauffman, Guiderdoni
\& White 1994; Cole~et~al.~1994; Baugh et al. 1997; Somerville \&
Primack 1998; Haiman \& Loeb 1997, hereafter HL97).  These
calculations have been quite successful in reproducing the observed
Tully--Fisher relation in the local Universe, the faint galaxy LFs out
to moderately high redshifts $z\lsim3$, and the recently discovered
clustering of Lyman--break galaxies at $z\sim3$ (Baugh et al. 1997;
Governato et al. 1998; Steidel~et~al.~1998).  Hierarchical models were
also applied to the more complex problem of quasar formation
(Efstathiou \& Rees 1988; Haehnelt \& Rees 1993; Small \& Blandford
1992), and were shown to reproduce the observed evolution in the
quasar LF at redshifts $z\ga 2$ under a minimal set of simplifying
assumptions (Haiman \& Loeb 1998b, hereafter HL98; see also Haehnelt,
Natarajan \& Rees 1998).

The above models were not tested in detail at redshifts much higher
than $z\sim3$ because of the lack of related observational data.  Such
tests are desirable from the theoretical point of view, since the
model predictions are expected to be more robust at higher redshifts
when the stars and quasars first condensed out of the well-defined set
of initial conditions in the early Universe. The exponentially low
abundance of collapsed objects at very early times reduces the
probability for complicated merger events.  Moreover, the chemical and
thermal conditions inside objects which assembled out of the pristine
intergalactic gas are better defined before metal enrichment and
thermal feedback from multiple generations of stars started to affect
the appearance of quasars and galaxies.  To date, the deepest optical
image of the Universe is the Hubble Deep Field (HDF,
Williams~et~al.~1996), which surveyed a $2.3^\prime\times2.3^\prime$
field in four bandpasses down to $\sim 29$ magnitude.  Although this
image has revealed a large number of galaxies of different
morphological types, a surprisingly large fraction of the sky was
found to remain dark at the HDF detection limit (e.g., Vogeley
1997). The Next Generation Space Telescope (NGST), scheduled for
launch in 2007, is planned to have a sensitivity of 1 nJy in the
1--3.5$\mu$m range, and should be able to detect globular cluster
scale objects out to $z\sim10$ (see Smith \& Koratkar 1998).

The properties of faint extended sources in the HDF (Madau et al.
1996) have been well described by detailed semi--analytic galaxy
formation models (Baugh et al. 1997).  On the other hand, the HDF has
revealed only a handful of faint unresolved sources (M\'endez et
al. 1996), and none with the colors expected for high redshift
objects.  The purpose of the present paper is to show that the lack of
point--like sources in the HDF can be used to constrain models for the
formation of cosmic structure at redshifts higher than previously
possible ($z\gsim4$).  The observational data and color--selection
criteria are summarized in \S~2 and \S~3, and the simplest versions of
the semi-analytic CDM models are outlined in \S~4. In \S~5, we compute
the optical and infrared quasar number counts in these models, and
show that the predictions from standard scenarios without a UV
feedback appear to be inconsistent with the observed lack of high--$z$
sources.  We also show that models with a UV feedback are marginally
consistent with the data.  In \S~6, we describe possible modifications
of these models that could reduce the predicted counts.  In \S~7, we
discuss the expected galaxy number counts. In \S~8, we present our
conclusions and argue that future infrared observations in the $J$ and
$H$ bands with NICMOS may in principle be used to place even stronger
constraints on this type of scenarios.

\section{High Redshift Quasar Candidates in the HDF}

Several surveys of unresolved objects in the HDF have appeared in the
literature (Elston, Santiago, \& Gilmore 1996; Flynn, Gould, \&
Bahcall 1996; M\'endez et al. 1996). M\'endez et al. (1996) were able
to identify 14 point--like objects, with only 6 of them at magnitudes
fainter than $V=26.5$\footnote{Magnitudes throughout this paper are in
the AB system, i.e.  ${\rm m_{X,AB}}=-2.5$ ${\rm
\log[f_{X}/erg~s^{-1}~cm^{-2}~Hz^{-1}]-48.6}$.}.  The faintest sources
in their sample tend to have rather blue $B-V$ and $V-I$ colors (see
also Elston et al. 1996). These authors point out that although the
star--galaxy separation software fails above $V\approx 27.5$,
misclassification at fainter magnitudes should turn more galaxies into
stars than vice versa, as the overwhelming majority of objects in the
HDF are galaxies and diffuse and faint features disappear first. Hence
the number of point--like sources detected down to $V=29$ should be
considered as an absolute upper limit to the number of stars or
quasars observed in this magnitude interval. In analogy with
star--forming galaxies, high redshifts ($z\gsim 3.5$) quasars are
expected to have very red $B-V$ and reddish $V-I$ colors because of
continuum absorption and line blanketing from neutral hydrogen along
the line of sight. As pointed out by Elston et al. (1996), however,
none of the stellar objects in the HDF has the colors expected for
high--$z$ quasars. A recent detailed search for faint quasars in the
HDF by Conti et al. (1998) confirms the lack of a population of very
red compact sources; no $z>3.5$ quasar candidates were found down to a
50\% completeness limit at $V=29$.

\section{Identifying Sources at ${\bf z\gsim 3.5}$}

\begin{figure}[t]
\includegraphics{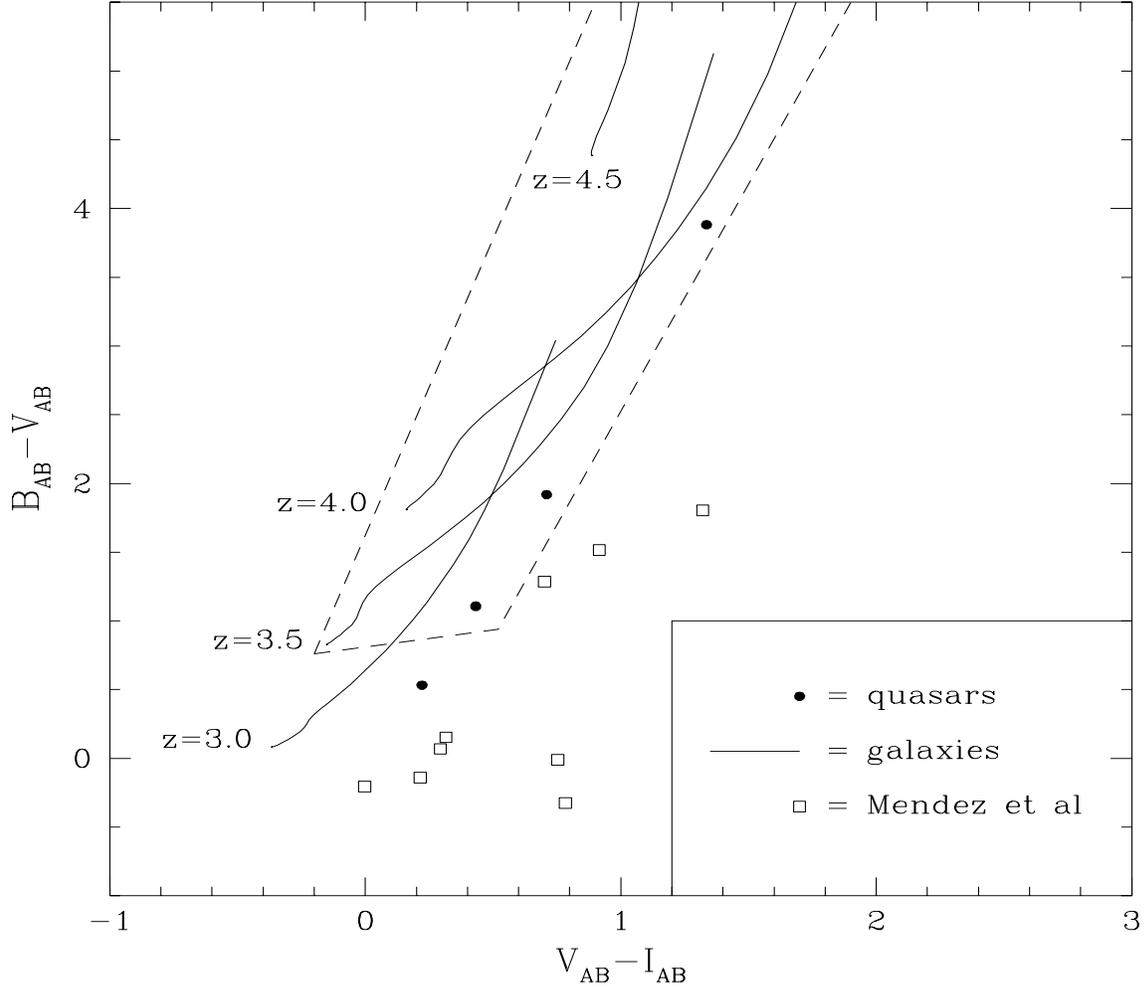}
\vspace*{4.5in}
\caption[$B-V$ vs. $V-I$ color--color diagram of $z>3$ objects]
{\label{fig:bvvi} Theoretical $B-V$ vs. $V-I$ color--color diagram for
$z>3$ objects.  The solid dots show quasars whose intrinsic colors are
assumed not to change with cosmic time, while the solid lines show
galaxies with ages of $0-10^9$ years. The data points ({\it empty
squares}) are the nine faint ($V>25$), unresolved objects from
M\'endez et al. (1996).  The trapezoidal region enclosed by the dashed
lines should selects all unreddened sources with $z\gsim 3.5$.}
\end{figure}

Before comparing our models directly to the observations, we show in
Figure~\ref{fig:bvvi} the location of the nine faint $V>25$ sources
detected by M\'endez et al. (1996) on a $B-V$ vs. $V-I$ color-color
diagram (empty squares).  The solid lines in this figure show the
predicted trajectories of model galaxies born at redshifts $3<z<4.5$
and passively evolving for $10^9$ years after an instantaneous burst
of star formation.  The filled dots show the expected colors of
quasars at the same redshifts.  Unlike galaxies -- which change their
colors due to the evolution of their stellar population -- we have
assumed that the aging of quasars only changes the overall
normalization, but not the shape of their intrinsic spectra (see \S 4
below for details of our model).  While all $z>3.5$ objects are
expected to fall within the trapezoidal region shown in the figure,
none of the M\'endez et al. (1996) sources appears to have the colors
predicted for a QSO or a galaxy at high--$z$.  One caveat to the
color--selection method is that some QSOs could have unusual colors.
Indeed, the three sources that have $(B-V)>1$ in the M\'endez sample
are found close to the border of the QSO selection region. However,
all of these objects are rather bright ($V$=25, 25.5, 26.3), and if
they are indeed QSOs, they are likely to have low ($z<3$) redshifts.
We note that if these three objects turned out to be QSOs at $z>3.5$,
then our models would underpredict the number of bright ($V\sim26$)
quasars, in addition to overpredicting the faint counts (see
Figure~\ref{fig:qv} and discussion below).

\begin{figure}[b]
\vspace{2.6cm}
\includegraphics{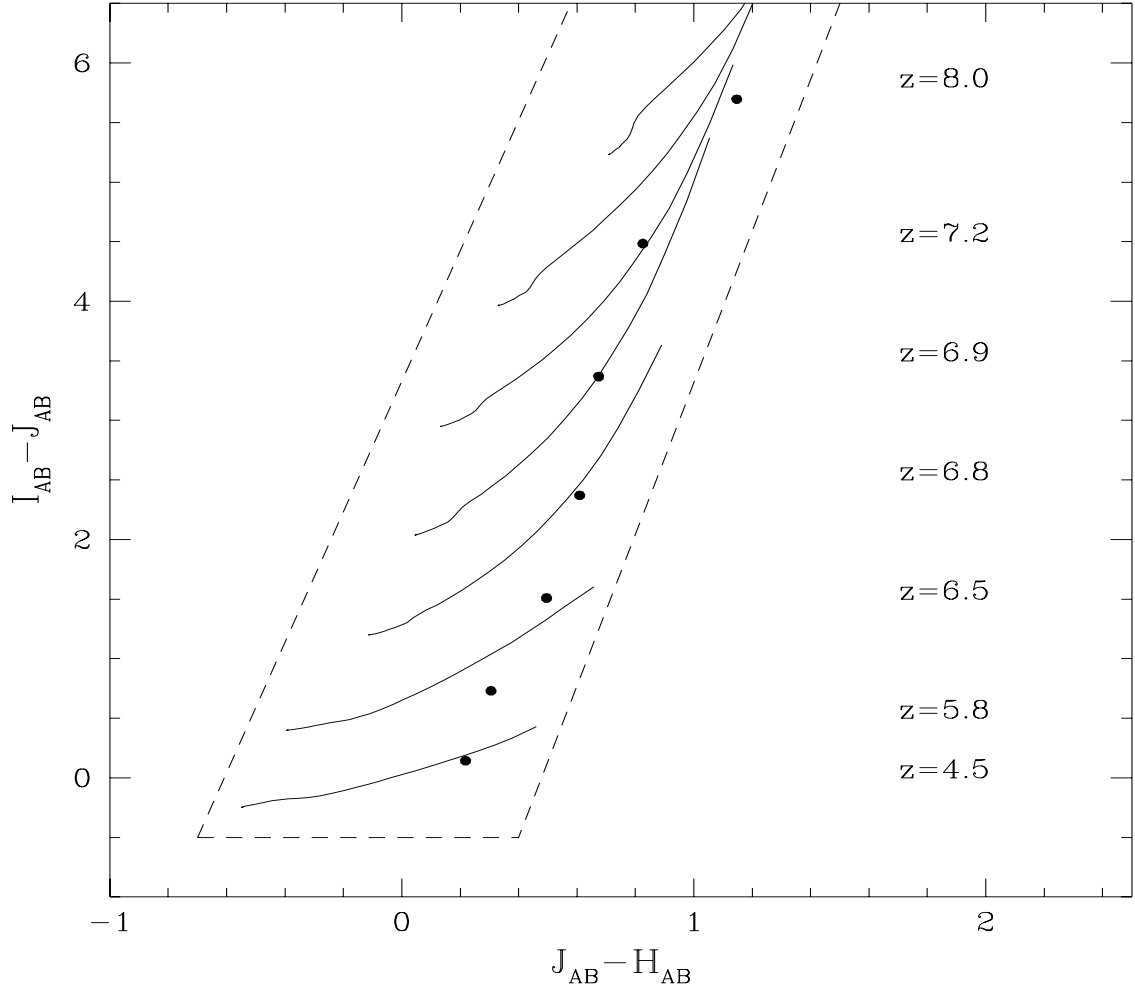}
\vspace*{4.5in}
\caption[$I-J$ vs. $J-H$ color--color diagram of $4.5\lsim z \lsim 8$
objects] {\label{fig:ijjh} Theoretical $I-J$ vs. $J-H$ color--color
diagram for $4.5\lsim z \lsim 8$ objects.  The solid dots show the
expected colors of quasars, while the solid lines show galaxies with
ages $0-10^9$ years.  The trapezoidal region selects objects with
$z\gsim 4.5$.}
\end{figure}

We note that it should be possible to identify objects at $z>4.5$ in
much the same way as the $z\sim 4$ galaxies are identified by the $B$
dropout technique.  Since at higher redshifts line blanketing and
Lyman--continuum absorption are even more pronounced due to the larger
number of intervening systems along the line of sight, a
straightforward extension of the dropout technique, using $I$, $J$,
and $H$ data, should work fairly well. In analogy with the $z\gsim
3.5$ $B$ dropout galaxies, we expect $V$, $I$, and $J$ dropout sources
to flag the redshifts $z\gsim$5, 7, and 9.  The $B-V$ vs.  $V-I$
diagram used to identify $z\sim4$ star-forming objects
(Madau~et~al.~1996) could then be replaced by a $I-J$ vs. $J-H$
color--color diagram as a probe of the redshift range $4.5<z<8$, as
illustrated in Figure~\ref{fig:ijjh}.  Unreddened galaxies and quasars
at $z>4.5$ are expected to be confined to the trapezoidal region
indicated in this figure, with negligible contamination from
foreground stars.

\section{Model Description}

In modeling the abundance of high--redshift sources, we assume that
the formation of dark matter halos follows the Press--Schechter (1974)
theory. Hence, the net rate of change in the comoving number density
of halos with mass $M_{\rm halo}$ is given by the derivative ${d\over
dz}(dn_{\rm c}/dM)$, where $dn_{\rm c}/dM$ is the Press--Schechter
mass function (comoving number density per unit mass).  This rate
includes a negative contribution from the disappearance of small halos
in merger events; the net ${d\over dz}(dn_{\rm c}/dM)$ becomes
negative when mergers dominate, i.e for masses below a characteristic
mass scale $M_*(z)$ at any given redshift $z$.  Quasars are
short--lived and we assume that their formation rate is proportional
to the formation rate of new halos. The negative contribution to this
rate from merger events should therefore be ignored. For relatively
bright quasars at high redshifts (and for all cases in our standard
$\Lambda$CDM model), the Press--Schechter formation rate is adequate
since the typical halo mass $M_{\rm halo}$ is above $M_*(z)$.
Otherwise, we make the conservative assumption that no quasars form in
halos with a mass $M_{\rm halo}<M_*(z)$ and therefore set ${d\over
dz}(dn_{\rm c}/dM)=0$ (see discussion below).

After each halo collapses and virializes, its gas might condense into
stars or a quasar black hole, provided it can cool efficiently from
its initial virial temperature.  Efficient cooling via collisional
excitation of atomic hydrogen requires the halo mass to be at least
$M_{\rm min}\sim10^{8}{\rm M_\odot}[(1+z)/11]^{-3/2}$ (HL97).  In
halos that exceed this mass, a fraction $\epsilon_\star$ of the gas
might fragment into stars -- forming a dwarf galaxy with a stellar
mass of $M_{\rm star}=\epsilon_\star M_{\rm halo}$.  A different
fraction $\epsilon_{\rm Q}$ of the gas might accrete onto a massive
central black hole with a mass $M_{\rm bh}=\epsilon_{\rm Q} M_{\rm
halo}$ and exhibit a low--luminosity quasar activity (Eisenstein \&
Loeb 1995; Loeb 1998).  The efficiencies $\epsilon_\star$ and
$\epsilon_{\rm Q}$ depend on several parameters, such as the redshift,
halo mass, or the initial angular momentum of the gas.  However, the
simplest assumption to make is that the efficiencies are the same for
all halos.

An important effect we need to consider is an external feedback from
photoionization, which exerts its influence by heating the gas before
it is able to cool and condense inside the dark matter potential
wells. Such a feedback is an inevitable consequence of the background
UV flux that builds up after the reionization epoch, when the
cosmological HII regions have overlapped and the universe is
transparent to the ionizing flux from each individual source.  Several
authors have discussed the consequences of this feedback (Efstathiou
1992; Quinn, Katz \& Efstathiou 1996; Thoul \& Weinberg 1996; Navarro
\& Steinmetz 1997). According to the spherically-symmetric simulations
of Thoul \& Weinberg (1996), a UV background at the level expected
from the proximity effect ($J_{21}=0.1-1$; Bajtlik, Duncan \& Ostriker
1988) completely suppresses the cooling and collapse of gas inside
halos with circular velocities $v_{\rm circ}\lsim 35~{\rm km~s^{-1}}$,
and substantially reduces (by $\approx 50\%$) the mass of cooled
baryons in systems with $v_{\rm circ}\la 50~{\rm km~s^{-1}}$.  While
this type of feedback has been mainly invoked to suppress the
formation of dwarf galaxies (Ikeuchi 1986; Efstathiou 1992; Babul \&
Rees 1992), it is also natural to expect that a similar mechanism may
inhibit the formation, or the fueling, of quasar black holes in
shallow potential wells.  In our standard model we therefore impose a
minimum circular velocity $v_{\rm circ}=50~{\rm km~s^{-1}}$ for the
halos of both luminous quasars and galaxies that form after the
reionization epoch.  This constraint results in a minimum halo mass of
$M_{\rm min}\sim10^{10}{\rm M_\odot}[(1+z)/11]^{-3/2}$, two order of
magnitudes higher than the cutoff mass obtained from the cooling
argument.

The value of $\epsilon_\star$ can be derived from the requirement that
the resulting stellar population reproduces the metallicity observed
at $z\sim 3$ in the Ly$\alpha$ forest, $10^{-2}\lsim Z_{\rm Ly\alpha}
\lsim 10^{-3}{\rm Z_\odot}$ (Tytler~et~al.~1995; Songaila \& Cowie
1996; Songaila 1997).  Assuming a Scalo (1986) stellar initial mass
function (IMF), and using carbon as a measure of metallicity, this
yields $0.017\lsim\epsilon_\star\lsim 0.17$ (HL97).  Similarly, the
value of $\epsilon_{\rm Q}$ for high--redshift quasars can be found by
matching the observed LF of quasars at intermediate redshifts,
$\phi(L,z)$.  Assuming that all quasars shine with a universal
light--curve in Eddington units [i.e. that $L(t)$ scales linearly with
the black--hole mass, $L(t)=M_{\rm bh} f(t) = \epsilon_{\rm Q} M_{\rm
halo} f(t)$], the constant $\epsilon_{\rm Q}$ and the function $f(t)$
can be adjusted so as to produce the observed $\phi(L,z)$ at $2.6\lsim
z\lsim 4$ based on the relation \beq \phi(L,z)=\int_{z}^{\infty}
\int_{M_{\rm min}}^{\infty} dM_{\rm halo} dz^{\prime} \frac{d^2n_{\rm
halo}}{dM_{\rm halo}dz^{\prime}}\delta[L-\epsilon_{\rm Q} M_{\rm halo}
f(t_{z,z^{\prime}})], \eeq where $t_{z,z^{\prime}}$ is the time
elapsed between the redshifts $z$ and $z^{\prime}$.  To account for
the decline of the observed quasar population at redshifts below
$z\sim2.6$ would require additional modeling, but should not effect
our conclusions here.  The fit to the observational data for
$\phi(L,z)$ (in the rest--frame $B$ band) given by Pei (1995) yields
$\epsilon_{\rm Q}\sim6\times10^{-4}$ and a light--curve shape that is
well approximated by an exponential, $f(t)\sim \exp(-t/t_0)$ with
$t_0=10^{5.8}~{\rm yr}$ in a flat $\Lambda$--CDM cosmology with
$\Omega_\Lambda=0.65$ and a Hubble constant $h=0.65$ (HL98).  Although
the best fit values for the $t_0$ and $\epsilon_{\rm Q}$ vary with
cosmology (see discussion in \S~6), the adopted procedure fixes
$\epsilon_{\rm Q}$ and $t_0$ to within a factor of two in any given
cosmological model (HL98).  We note that $t_0$ could be increased
substantially if black holes formed in only a fraction $f$ of all
halos.  For example, Eisenstein \& Loeb (1995) have argued that only
the small subset of initial seed perturbations with sufficiently slow
initial rotation can form black holes.  In order to recover the fit to
the luminosity function, the lifetime would then need to be increased
to $\sim t_0/f$.

The Press--Schechter halo formation rate depends on the cosmology and
power spectrum.  For our ``standard'' scenario we have adopted the
concordance model of Ostriker \& Steinhardt (1995), i.e. a flat
$\Lambda$CDM model with a slightly tilted power spectrum
($\Omega_0,\Omega_\Lambda, \Omega_{\rm
b},h,\sigma_{8h^{-1}},n$)=(0.35, 0.65, 0.04, 0.65, 0.87, 0.96).
Convenient expressions for the differential volume element, luminosity
distance, and time--redshift relation in this model were given in
terms of elliptic integrals by Eisenstein (1997); an accurate fitting
formula for the growth function is given by Carroll~et~al.~(1992).  We
also consider two open models without a cosmological constant, one
with $\Omega_0$=0.35 and the other with $\Omega_0=0.2$.  In order to
remain consistent with the power--spectrum normalization on both COBE
and cluster scales (Viana \& Liddle 1996), we change the spectral
index to $n$=1.15 and 1.2 respectively in these models.

For our quasar template spectrum, we have used the average spectrum
derived from the observations of 47 quasars by
Elvis~et~al.~(1994). The stellar template spectra have been computed
from the low--metallicity ($Z=4\times 10^{-4}$) population synthesis
models of Bruzual \& Charlot (1996), assuming an instantaneous burst
of star formation and a Scalo (1986) IMF.  Note that a Salpeter IMF
produces about twice as many UV photons as the Scalo IMF, for the same
level of carbon enrichment of the intergalactic medium (IGM).  We find
that adopting a Salpeter IMF would increase our predicted number
counts down to $V\sim 30$.  In our standard model, the increase is
negligible at $V=30$ but is an order of magnitude at $V=22$.

\section{Number Counts}

In both the quasar and galaxy models, the number of sources
located between redshifts $z_{\rm min}$ and $z_{\rm max}$ per unit
solid angle, with observed visible magnitude between $V$ and $V+dV$,
is given by
\beq 
\frac{dN}{d\Omega dV}(V,z_{\rm min},z_{\rm max})=
\int_{z_{\rm min}}^{z_{\rm max}} dz 
\left(\frac{dV_{\rm c}}{dzd\Omega}\right) n_{\rm c}(z,V),
\eeq
where $dV_{\rm c}/dzd\Omega$ is the comoving volume element per unit
redshift per unit solid angle (Eisenstein 1997), and $n_{\rm c}(z,V)=
dN/dVdV_{\rm c}$ is the comoving number density of objects at a
redshift $z$ with observed magnitudes between $V$ and $V+dV$.

Based on the halo formation rate $d^2n/dMdz$, $n_{\rm c}$ is given
by a sum over halos of different ages that exist at each redshift,
\beq 
n_{\rm c}(z,V)=\int_z^{\infty} dz^\prime
\frac{dM_{\rm halo}}{dV}(z,z^{\prime},V) \left.\frac{d^2n_{\rm c}}{dM dz^\prime} 
\right|_{M_{\rm halo}(z,z^{\prime},V),z^\prime},
\label{eq:ncom} 
\eeq
where the factor $dM_{\rm halo}/dV$ converts the number density per unit
mass interval to number density per unit flux interval.  The $V$ magnitude
in the AB system is given by
\beq
V=-2.5\log f_V-48.60,
\eeq
where the observed $V$--band flux $f_V(z,z^\prime,M_{\rm halo})$ from a
halo at redshift $z$ with a mass $M_{\rm halo}$ that has formed at
$z^\prime\ge z$ equals (in units of ${\rm erg~sec^{-1}~Hz^{-1}~cm^{-2}}$),
\beq
f_V(z,z^\prime,M_{\rm halo})=
\frac{(1+z)\epsilon M_{\rm halo}}{4\pi d_{\rm L}(z)^2}
\left[\int_{0}^{\infty} d\nu T_V(\nu)
\exp\left[-\tau(\nu,z)\right] 
j(\nu(1+z),t_{z,z^\prime}) \right]\times
\left[\int_{0}^{\infty} d\nu T_V(\nu) \right]^{-1}.
\label{eq:halomag}
\eeq Here $d_{\rm L}(z)$ is the luminosity distance;
$\epsilon=\epsilon_{\rm Q}\equiv6\times 10^{-4}$ or
$\epsilon=\epsilon_\star\equiv (\Omega_{\rm b}/\Omega_0) \times
(0.017-0.17) =0.002-0.02$ is the fraction of halo mass in the form of
the central black hole or stars, respectively; $\tau(\nu,z)$ is the
effective optical depth associated with intervening Ly$\alpha$ forest
clouds and Lyman--limit systems between redshifts $0$ and $z$ at the
observed frequency $\nu$ (Madau 1995); $T_V(\nu)$ is the
frequency--dependent transmission of the $V_{606}$
filter\footnote{Available at
http://www.stsci.edu/ftp/instrument\_news/WFPC2/Wfpc2\_faq/wfpc2\_phot\_faq.html.};
and $j(\nu,t_{z,z^\prime})$ is the template luminosity per unit
(stellar or black hole) mass for either stars or quasars, in units of
${\rm erg~sec^{-1}~Hz^{-1}~M_\odot^{-1}}$.  The time dependence of
$j(\nu)$ is obtained from our fitted light--curve function $L(t)$ for
quasars, and from tabulated evolutionary tracks for stars
(Schaller~et~al.~1992; see HL97 for details).  We suppressed the
emergent flux from all objects at emission frequencies above
\lya~before reionization occurs (assuming sudden reionization at the
redshift $z=13$ or $9$ for stars, and $z=11.5$ for quasars; see HL98
for details), since the Gunn--Peterson optical depth of HI at these
frequencies is exceedingly high prior to reionization.

The expected counts of $z>3.5$ quasars in the $V$ band are shown by
the solid lines, labeled A, in Figure~\ref{fig:qv}.  The bottom panel
also shows the observational data points from the $B$ band (Pei 1995),
converted to $V$ counts using the shape of our template spectrum.  By
construction, the differential model number counts for quasars in the
narrow redshift interval $3<z<3.2$ are consistent with the data
between $17<V<22.5$.  The upper panel in this figure shows the
cumulative counts of $z>3.5$ quasars, and reveals that down to the
$V=29$ limit of the HDF our standard model predicts about $4$ objects.
Since no unresolved sources at $z>3.5$ have been found in the HDF, a
model which is consistent with the HDF data at the $2\sigma$ level
should predict no more than $2$ such high-$z$ QSOs.  On the other
hand, since the observations are only $\sim50$\% complete, a
prediction of 4 QSOs could be allowed. Our standard quasar scenario
therefore appears to be marginally consistent with the HDF data.
Although the existing data on the LF only constrains the individual
quasar model parameters such as $\epsilon_{\rm Q}$ and $t_0$ to within
a factor of $\sim2$, the $z>3.5$ number counts are predicted to a much
better accuracy, since the model is calibrated by simultaneously
adjusting $\epsilon_{\rm Q}$ and $t_0$ to reproduce the quasar LF at
$z\sim 3$.  As a result, an overprediction of the quasar counts by a
factor of $2$ could still be significant.  It is important to remember
here that in our standard model, we have eliminated quasars in halos
with $v_{\rm circ}<50~{\rm km~s^{-1}}$ to indicate the effect of
photoionization heating.  The thin dashed curves in
Figure~\ref{fig:qv}, labeled A1, show the number counts in a model
without this UV feedback, and reveal a prediction of $\sim 8$ objects
to $V=29$.  The overprediction in this model would be rather severe, a
factor of $\sim 4$.

The expected number counts of $z>3.5$ galaxies in our standard model
are shown by the curves labeled A in Figure~\ref{fig:sv}. The solid
and dashed lines bracket the expected metallicity range of the IGM at
$z\approx 3$, ${\rm 10^{-3}Z_\odot< Z<10^{-2}Z_\odot}$.  Depending on
the metallicity calibration, the total number of galaxies observable
down to $V=29$ mag is between 20 and 140.  This exceeds the predicted
quasar counts by a factor of $5-35$, and is $10-100$ larger than what
would be allowed for point--sources in the HDF.  However, depending on
their assumed intrinsic sizes, some or most of the $z>3.5$ galaxies
could actually be resolved (see discussion below). We also note that
recent revisions of the C/H ratio in these absorbers prefer lower
metallicities, closer to $10^{-3}{\rm Z_\odot}$ (Songaila 1997).

\section{Reducing the ${\bf z>3.5}$ Quasar Counts}

As the simplest scenarios with a UV feedback appear to be only
marginally consistent with the faint $V$ counts of quasars, the
question naturally arises: how should one modify the models in order
to reduce even further the expected number of faint point--like
objects?  In general, the theoretical number counts are sensitive to
the underlying cosmology and power spectrum, the relation between the
halo mass and the emitted luminosity, as well as the time--evolution
of each object, including its merger history.

\begin{figure}[b]
\vspace{2.6cm} 
\includegraphics{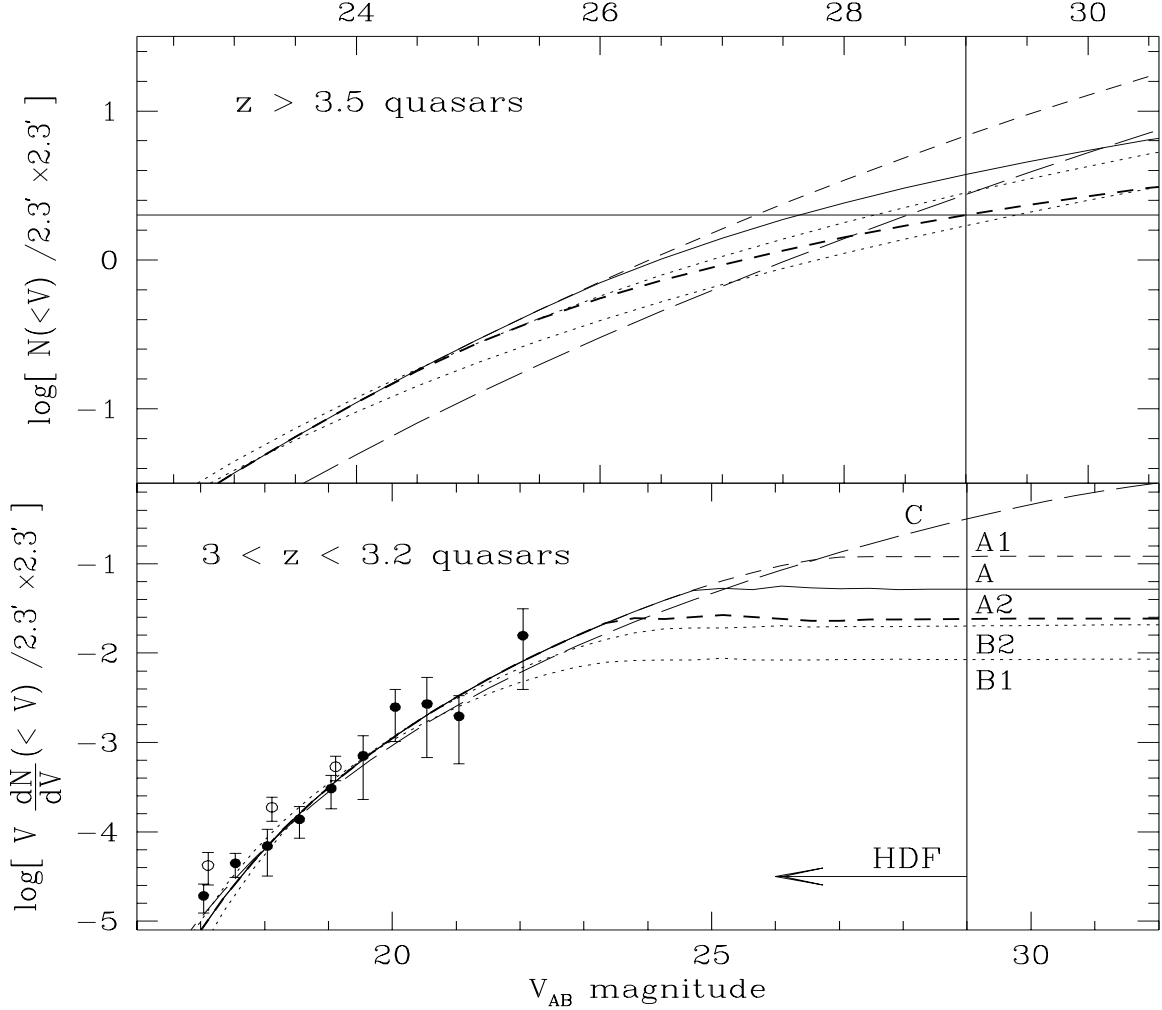}
\vspace*{4.5in}
\caption[Number counts for high--redshift quasars in the $V$ band]
{\label{fig:qv} $V$-band number counts for high--redshift quasars. The
lower panel shows differential counts for $3<z<3.2$, as well as data
adapted from Pei (1995).  The upper panel shows cumulative number
counts for $z>3.5$.  The curves labeled A show predictions of our
standard $\Lambda$CDM model with ${\rm M_{bh}\propto M_{halo}}$.  The
curves labeled A1 show the counts in a model with no UV feedback. The
curves labeled A2 correspond to a model with halo circular velocities
$v_{\rm circ} \geq 75~{\rm km~s^{-1}}$.  The curves labeled B1 and B2
correspond to open CDM models with $\Omega_0$=0.35 and 0.2,
respectively, and the curves labeled C corresponds to a ``tilted''
model with ${\rm M_{bh}\propto M_{halo}^{1.5}}$.}
\end{figure}

\begin{figure}[b]
\vspace{2.6cm}
\includegraphics{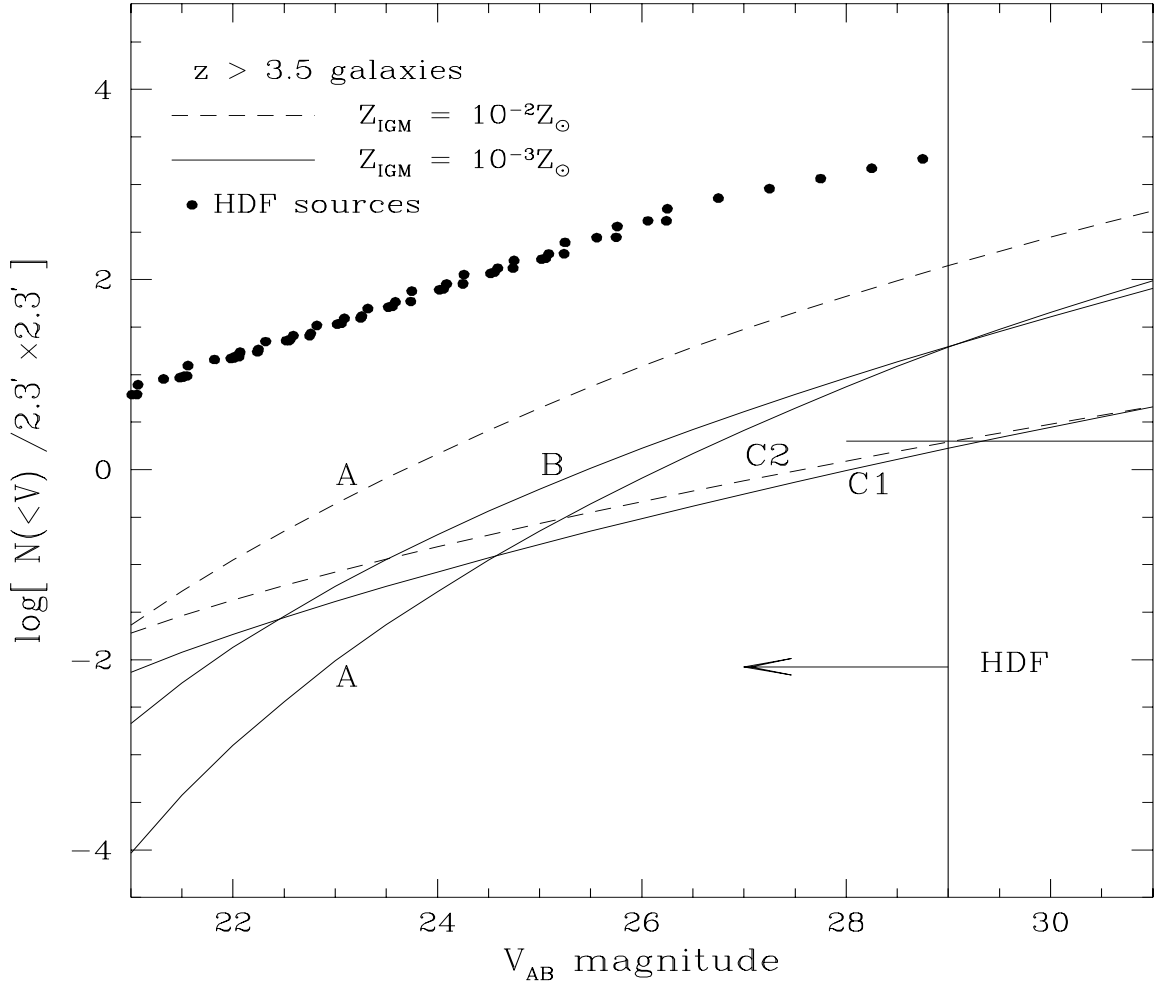}
\vspace*{4.5in}
\caption[Number counts for high--redshift galaxies in the $V$ band]
{\label{fig:sv} $V$ counts for high--redshift galaxies in the simplest
model with $M_{\rm star}\propto M_{\rm halo}$.  The solid lines
correspond to low star formation efficiencies ($M_{\rm star}=0.017
M_{\rm halo}$) that produce an IGM metallicity of $10^{-3}Z_\odot$,
and the dashed lines correspond to a star formation efficiency which
is 10 times higher.  The curves labeled A correspond to our standard
model. The curve labeled B shows the counts in the open CDM model. The
curves labeled C1 and C2 correspond to models with ${\rm
M_{star}\propto M_{halo}^3}$ (for $Z_{\rm Ly\alpha}=10^{-3}{\rm
Z_\odot}$) and ${\rm M_{star}\propto M_{halo}^{3.5}}$ (for $Z_{\rm
Ly\alpha}=10^{-2}{\rm Z_\odot}$), respectively.  The solid dots show
all extended sources detected in the HDF in the $V$ band (Williams et
al. 1996).}
\end{figure}

The simplest explanation for the lack of faint quasars would be a more
effective UV feedback mechanism, resulting in a higher minimum halo
size for luminous objects.  In Figure~\ref{fig:qv}, the lines labeled
A2 show the number counts of quasars in a model with $v_{\rm
circ}>75~{\rm km~s^{-1}}$.  The top panel of this figure demonstrates
that with this cutoff value of $v_{\rm circ}$, the $z>3.5$ counts are
reduced to 2 objects, i.e. consistent with the observed lack of
detections.  This required cutoff value is higher than the values
obtained from spherical simulations (Thoul \& Weinberg 1996). With a
UV background at the level of $0.3<J_{21}<1$, the cooled baryon
fraction is fully suppressed for halos with $v_{\rm circ}\lsim 35~{\rm
km~s^{-1}}$, suppressed by 50\% for $v_{\rm circ}\lsim 50~{\rm
km~s^{-1}}$, and unaffected for circular velocities above $v_{\rm
circ}\gsim 75~{\rm km~s^{-1}}$.  However, using 3--D simulations,
Navarro \& Steinmetz (1997) find the UV feedback to be more effective,
and their circular velocity thresholds are $v_{\rm circ}=80-200~{\rm
km~s^{-1}}$, exceeding the required value $v_{\rm circ}=75~{\rm
km~s^{-1}}$ derived here.

In the absence of a photoionization feedback, another straightforward
approach to reduce the faint counts is to make the slope of the
luminosity function shallower.  If the luminosity of each object were
steady and proportional to the halo mass, $L\propto M$, then the LF
would have been related to the mass function simply by
$\phi(L)=dn_{\rm c}/dM\times dM/dL\propto dn_{\rm c}/dM$, i.e. the LF
would have had the same slope as the Press--Schechter mass function.
At low masses $dn_{\rm c}/dM\propto M^{-2+(n-1)/6}$, independent of
cosmology and only very weakly dependent on the power spectrum index
$n$.  By introducing a non--linear scaling between luminosity and halo
mass, $L\propto M^{1+\alpha}$ (see discussion below) one could tune
the LF to have a shallower slope, $\phi(L)\propto
L^{-1-1/(1+\alpha)}$.

In reality, each source shines only for a limited amount of time
$\Delta t$ which is much shorter than the Hubble time.  This results
in $\phi(L)\sim \Delta t\times dn_{\rm c}/dMdt\times dM/dL(t)$, where
$dM/dL\propto \exp(-t/t_0)$ for our exponential light--curve.  Here, a
complication arises because at masses below $M_*$, the time derivative
of the mass function is negative, $dn_{\rm c}/dMdt<0$.  Physically,
this derivative is the sum of a positive term (the formation rate of
new halos with mass $M$), and a negative term (the disappearance rate
of existing halos due to mergers), and while the formation rate
exceeds the merger rate for $M>M_*$, the latter dominates for $M<M_*$.
The exact ratio between the formation and merger rates requires a
scheme that goes beyond the Press--Schechter theory and describes the
merging history of halos (Lacey \& Cole 1993; Kauffmann \& White
1993).  For the sake of simplicity, we will assume here that no halos
form below the threshold mass $M_*$ and that no halos are destroyed by
mergers above this mass, i.e. we simply set the derivative $dn_{\rm
c}/dMdt$ to zero for $M<M_*$ and to the Press-Schechter value for
$M>M_*$.  Although this assumption is ad--hoc, it can only
underestimate the number of objects.  Therefore, the reduction in the
number counts derived here should be considered as the minimum amount
of change that is necessary in order to fit the observations.

Under the above prescription, the LF flattens to $\phi(L)\propto 1/L$
at luminosities corresponding to $M<M_*$.  The cosmological parameters
(${\rm \Omega_0, \Lambda, H_0}$) and the power spectrum parameters
($\sigma_{\rm 8h^{-1}},n$), flatten the number counts at $V=29$ (with
$VdN/dV$=const) in models where low--mass halos form at sufficiently
high redshift.  Successful models raise the threshold mass $M_*$ at
$3\lsim z\lsim6$ to high enough values so that $V\sim29$ objects
become associated with sub--$M_*$ halos.  For example, at $z\sim 3$,
the halo mass corresponding to $V\sim29$ in our models is $\sim
10^{8-9} M_\odot$.  In general, $M_*$ is high in models with large
values of $\sigma_{\rm 8h^{-1}}$, $n$, and $\Omega_0$, and low values
of $\Lambda$.  These parameters are related by the normalization of
the power spectrum on the COBE (X--ray cluster) scale, which requires
$\sigma_{\rm 8h^{-1}}$ to decrease (increase) with decreasing
$\Omega_0$.  In Figure~\ref{fig:qv}, we show two $\Lambda=0$ open
models, in which $\Omega_0=0.35$ and 0.2 (lines labeled B1 and B2,
respectively).  The values of ($\sigma_{\rm 8h^{-1}}$ and $n$) in
these models which are consistent with both the COBE and cluster
normalizations are (0.87, 1.15) and (0.87, 1.2), respectively.  Note
that the values of $t_0$ and $\epsilon_{\rm Q}$ also need to be
readjusted in these cosmologies. We find ($t_0,\epsilon_{\rm
Q}$)=($10^{6.2}$yr, $10^{-3.7}$) for $\Omega_0=0.35$ and
($t_0,\epsilon_{\rm Q}$)=($10^{6.8}$yr, $10^{-3.9}$) for
$\Omega_0=0.2$.  The lines in the bottom panel labeled B1 and B2 show
that the $z\sim3$ number counts indeed flattens at $V\sim23$ and
$V\sim24$ in these models.  The upper panel of this figure also shows
that in the $\Omega_0=0.35$ model, the cumulative $z>3.5$ counts at
$V=29$ are lowered to the required level of $\sim 2$ objects. While
the $z>3.5$ counts in the $\Omega_0=0.2$ model are also reduced, the
predicted number of quasars at the HDF limit is still too large in
this case (i.e. $\sim3$ objects, making the observed lack of
detections a $3\sigma$ event).

A third alternative approach to the photoionization feedback for
reducing the faint $z>3.5$ counts is to make the LF decline faster at
high redshifts.  While the quasar LF is reasonably well determined at
$z\sim3$, it is significantly more uncertain around $z\sim4$ (see
Figs. 1 and 2 in Pei 1995).  One could exploit this uncertainty and
modify the models so that the predicted number of $z\sim4$ objects is
at the low end of the allowed range.  We have found that the redshift
evolution can not be made sufficiently steep by changing the cosmology
or the power spectrum, while still maintaining a good fit to the LF
around $z\sim3$.  However, the evolution can be made steeper if the
halo mass associated with a fixed luminosity is increased, since the
formation rate of higher mass halos evolves more rapidly.  To achieve
this change, we introduce a simple dependence of luminosity on halo
mass, $L=\epsilon_0(M_{\rm halo}/10^8{\rm M_\odot})^{1+\alpha}$ with
$\alpha>0$.  A similar scaling, including an additional redshift
dependence, $L\propto M_{\rm bh}\propto v_{\rm circ}^5\propto M_{\rm
halo}^{5/3}(1+z)^{5/2}$ has been recently considered by Haehnelt,
Natarajan \& Rees (1998).  Our adopted scaling lacks the factor
$(1+z)^{5/2}$, and therefore predicts a steeper redshift evolution and
a slightly smaller number of $z>3.5$ objects than in Haehnelt et
al. (1998).  Since most objects at $V=29$ are located at $z<6$, the
difference in the number counts predicted by the two models in the
relevant magnitude range is small ($\lsim15\%$).

For any given value of $\alpha$, the fitting procedures described in
\S~4 can be repeated to obtain the values of $\epsilon_0$ and $t_0$.
Although by definition our procedure for quasars fits the LF at
$2.6<z<4$, an increase in the value of $\alpha$ yields an increase in
the characteristic mass of the LF, and a decrease in the total number
of faint $z>3.5$ quasars.  We have experimented with values of
$0<\alpha<1$, and found that for $\alpha \gsim 0.5$, the fitting
procedure for quasars fails, due to the implausibly steep redshift
evolution of the model LF already for the observed redshift interval
$2.6<z<4$.  In other words, an acceptable fit in our models to the LF
at $z=3$ would result in a predicted $z\sim4$ LF that falls below the
errorbars of the observed $z=4$ bin in Pei (1995).  Note that much
stronger deviations from the linear relation $M_{\rm bh}\propto M_{\rm
halo}$ would also be mildly inconsistent with observational data for
local galaxies (Magorrian~et~al.~1997; McLeod 1997).  The largest
plausible reduction in the faint counts can therefore be achieved
using $\alpha=0.5$.  In this case, we obtain $M_{\rm bh}/M_{\rm
halo}=8\times10^{-7}(M_{\rm halo}/{\rm 10^8M_\odot})^{0.5}$ and
$f(t)=\exp(-t/3\times10^6 {\rm yr})$, and predict the number counts
shown by the curves labeled C in Figure~\ref{fig:qv}.  The bottom
panel shows that the quality of the fit to the $z\sim 3$ LF at
$17<V<22.5$ is comparable to the fit in our standard model.  The top
panel shows that relative to model A, the $z>3.5$ counts decrease by a
factor of 3 to $\sim 3$ objects. However, even with the largest
allowed value of $\alpha$, the $z>3.5$ counts remain a too high
(i.e. $\sim3$ objects, making the HDF a $3\sigma$ event).

A fourth effect that could contribute to the deficit of observed faint
quasars is the expected shift in the ``big blue bump'' (BBB) component
of the quasar spectrum to higher energies for the lower mass black
holes which populate the Universe at high redshifts.  In thin
accretion disk models (e.g., Peterson 1997) the emission temperature,
and hence the peak wavelength of the BBB, scales with the black hole
mass as $\lambda_{\rm peak}\propto M_{\rm bh}^{0.25}$.  In our present
template spectrum, which is based on data from local ($z\lsim 1$)
quasars (Elvis et al. 1994), the BBB peak is tentatively located at
$1000$\AA~(see HL98). Since the characteristic black hole might be
smaller by several orders of magnitude at $z\sim4$--$6$ than at
$z\lsim 1$, the BBB feature might shift by a factor of a few in
wavelength.  Such a shift would reduce the observed $V$ band flux by a
factor of $\sim7$, and the rest--frame $B$ band flux (which we use to
calibrate our model) by a factor of $\sim 2$.  In particular, if the
BBB in high--redshift quasars is blueshifted to $\la 300$\AA, then the
faint $V$ quasar counts would be reduced by a factor of $\sim$2, not
quite sufficient to bring the number counts into agreement with the
HDF data.

\section{Number Counts of ${\bf z>3.5}$ Galaxies}

The predicted number of $z>3.5$ galaxies in our models at the HDF
limits is a factor of 5--35 higher than the number of quasars.
However, faint galaxies might have sufficiently large angular sizes so
as to appear extended and avoid the overprediction of point sources.

A simple estimate shows that this might indeed be the case. To
estimate what fraction of the $z>3.5$ galaxies may have angular
diameters above the 0.1\H{} pixel size of the three wide field cameras
(WFC), we assume that each collapsed halo has a $1/r^2$ density
profile, with an overdensity of $50$ at the edge of the halo relative
to the average background density (equivalent to an average
overdensity of 150), and that the luminous region of the galaxy
extends out to 5\% of the virial radius of the halo (see Navarro,
Frenk \& White 1995).  With these assumptions, we find that for the
low star formation efficiency (producing an average metallicity of
$Z_{\rm Ly\alpha}=10^{-3}{\rm Z_\odot}$ in the intergalactic medium),
only 0.5\% of the $z>3.5$ objects would be unresolved at $V=29.8$.  As
a result, the HDF is expected to contain no point--like galaxies at
this magnitude limit.  The unresolved fractions around the HDF
sensitivity limit are significantly higher for a high star formation
efficiency ($Z_{\rm Ly\alpha}=10^{-2}{\rm Z_\odot}$).  In this case, a
fraction equal to 2\%, 9\%, 18\%, and 24\% of the $z>3.5$ objects
would be unresolved at $V=27.5$, $28.0$, $28.9$ and $29.8$,
respectively.  Therefore, given the expected intrinsic sizes of the
galaxies, the overprediction problem disappears at low star formation
efficiencies, but is not fully resolved if the star formation
efficiency is higher and the apparent angular size of galaxies
smaller.  This could be interpreted as support for a low star
formation efficiency, and a corresponding low average metallicity of
the IGM at $z\sim 3$.

It is interesting to examine the sensitivity of the number counts to
the underlying cosmology and to the relation between $M_{\rm star}$
and $M_{\rm halo}$ assuming that the stars occupy a region much
smaller than 5\% of the virial radius of each halo so that most
$z>3.5$ galaxies are unresolved.  Obviously, it is more difficult to
reconcile the number counts of such galaxies with the HDF data than
that of quasars.  There are two reasons for this result. First, the
number counts of galaxies are higher to begin with.  Second, the model
galaxies are $\sim2$ orders of magnitude less luminous than quasars,
so that the same $V$ magnitude corresponds to a halo mass which is
larger by $\sim 2$ orders of magnitude.  The differential number
counts flatten at the fixed mass $M_*$; for galaxies this mass
corresponds to $V\sim30$ at $z=4.5$.

As an example, the curve labeled B in Figure~\ref{fig:sv} shows the
galaxy counts in the $\Omega_0=0.35$ open CDM model.  The counts are
somewhat less steep than in the $\Lambda$ model (the solid curve
labeled A).  Although the flattening of the mass function at $M<M_*$
reduces the faint counts, the effect is negligible at $V\sim29$.  The
figure also shows the results of changing the dependence of the star
formation efficiency on the halo mass (changing $\alpha$).  When the
star formation efficiency is normalized to $Z_{\rm Ly\alpha}=10^{-3}
{\rm Z_\odot}$, then $\alpha=2$ and $\epsilon_0=3.3\times10^{-9}$ are
required for consistency with no detections (see the line labeled C2).
Similarly, when $Z_{\rm Ly\alpha}=10^{-2} {\rm Z_\odot}$ is assumed,
$\alpha=2.5 $ and $\epsilon_0=7.3\times10^{-12}$ are required (C1).
As the figure shows, the slopes of the latter number count curves are
flattened considerably.  For reference, we show also the number of
counts of all extended sources detected in the HDF in the $V$ band.
These counts have a shallow slope and demonstrate that at the $V=29$
limit only a small fraction of all detected objects are expected to be
at $z>3.5$.  Finally, we note that an additional reduction of the
galaxy counts could also be caused by dust extinction.  Estimates
based on the Calzetti~et~al.~(1994) extinction curves indicate that
the number of $z\lsim 3$ sources could be as much as $\sim7$ times
higher than currently detected (Pettini~et~al.~1997), when dust
reddening is taken into account.

Given all of the above uncertainties, we conclude that, unlike in the
case of quasars, there is no significant discrepancy between the
expected galaxy counts and the lack of faint point sources.  This is
true as long as the stars occupy a region extending out to a few
percent of the virial radius of their host halo, and the
star--formation efficiency is close to the value expected from a low
average Ly$\alpha$ forest metallicity, $Z_{\rm Ly\alpha}=10^{-3}{\rm
Z_\odot}$.

\section{Conclusions}

The observed lack of faint, unresolved, and red sources in the HDF
down to $V\approx 29$ requires models that are consistent with the
data at the $2\sigma$ level to predict no more than $2$ point--like
high-$z$ objects to this limiting magnitude. Given that the
observations are $\sim50$\% complete, this appears to be marginally
consistent with the predicted number ($\sim4$) of high redshift
quasars in the simplest hierarchical models, which suppress the
formation of black holes in CDM halos with circular velocities $v_{\rm
circ} \leq 50~{\rm km~s^{-1}}$.  Without a photoionization feedback,
these models predict $\sim 8$ quasars from $z>3.5$. In models without
a UV feedback, the predicted quasar counts might be reduced by early
structure formation (so that $M_* \gsim 10^{11}{\rm M_\odot}$ by
$z=3$), but only if one postulates that new black holes do not form in
halos below the nonlinear mass--scale, $M_*(z)$. An example for such a
model is the $\Omega_0=0.35$ open CDM model, with a power spectrum
slope $n=1.15$ and normalization $\sigma_{8h^{-1}}=0.87$.  The $z>3.5$
number counts can be further reduced by introducing a black hole
formation efficiency $M_{\rm bh}/M_{\rm halo}$ that depends on the
halo mass as $\propto M_{\rm halo}^{0.5}$. Similarly, a shift of the
big blue bump component to shorter wavelengths in the spectra of faint
quasars could also help reduce the predicted counts.  Combined with a
minimum circular velocity of $v_{\rm circ}= 50~{\rm km~s^{-1}}$,
either of these latter two effects could reduce the number counts to
less than $\sim 2$ objects.  On the other hand, we find that without a
photoionization feedback, neither of these effects would be sufficient
by itself to explain the lack of detections.

The models predict $\sim$ 5--35 times more $z>3.5$ galaxies than
quasars at $V=29$. If the star formation rates in these galaxies are
adjusted to produce an average Ly$\alpha$ forest metallicity of
$10^{-2}{\rm Z_\odot}$ by $z=3$, then more than $10\%$ of all galaxies
could appear unresolved at the 0.1\H{} pixel size of the WFC, possibly
leading to another discrepancy with the observed lack of detections.
The situation changes if the star formation rate is lowered by an
order of magnitude (normalized to produce a $10^{-3}{\rm Z_\odot}$
average IGM metallicity at $z\sim 3$), since then almost all of the
predicted $z>3.5$ galaxies are expected to be extended rather than
point sources.

\begin{figure}[b]
\vspace{2.6cm}
\includegraphics{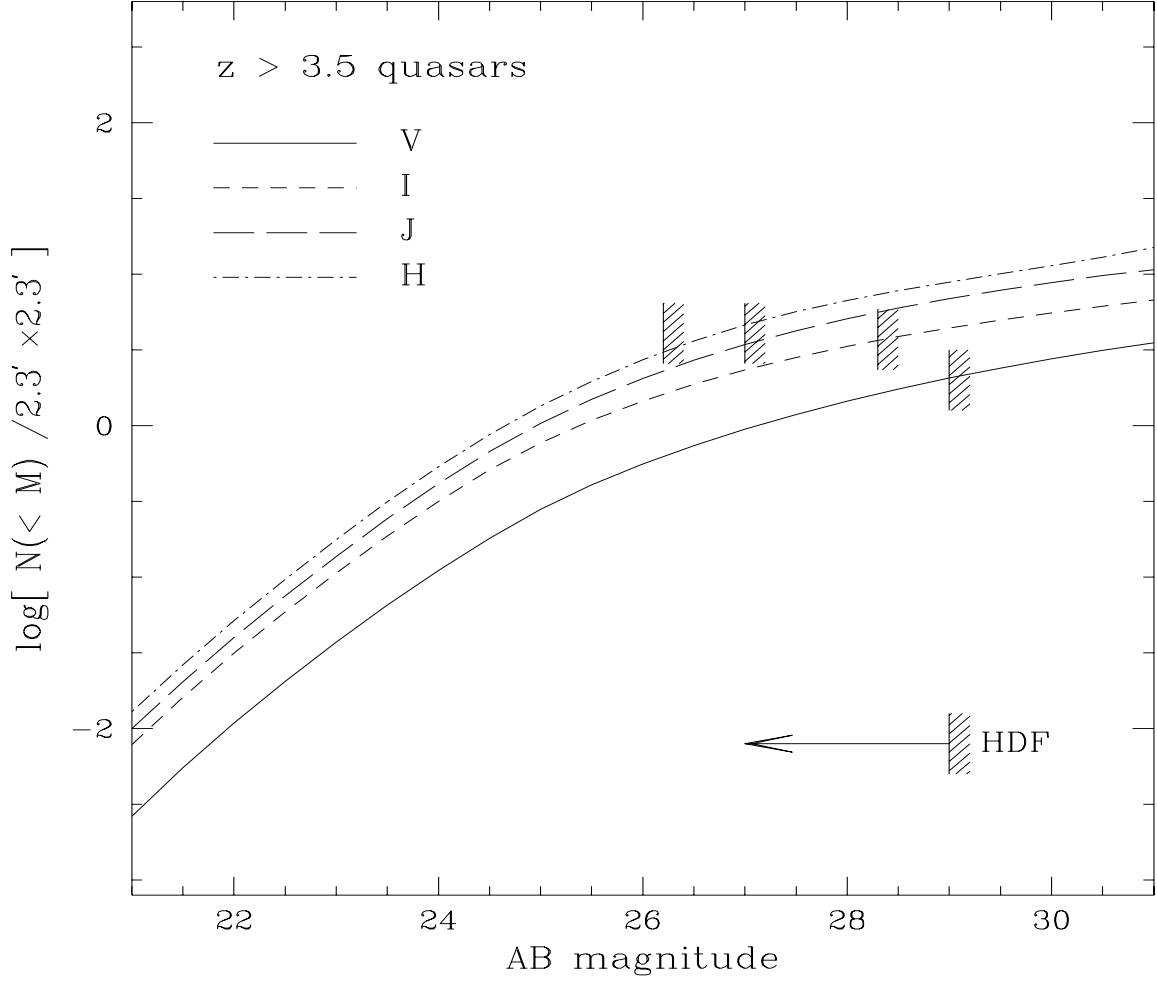}
\vspace*{4.5in}
\caption[$V$, $I$, $J$, and $H$ counts for galaxies]
{\label{fig:qvijh} $V$, $I$, $J$, and $H$ counts for quasars in the
$\Lambda$CDM model with $\Omega_0$=0.35 and minimum circular velocity
$v_{\rm circ}= 75~{\rm km~s^{-1}}$.  This model is consistent with HDF
data in the $V$ band at the 5\% level.  Sensitivities shown for each
band are for a fixed signal--to--noise ratio and exposure time.}
\end{figure}

\begin{figure}[b]
\vspace{2.6cm}
\includegraphics{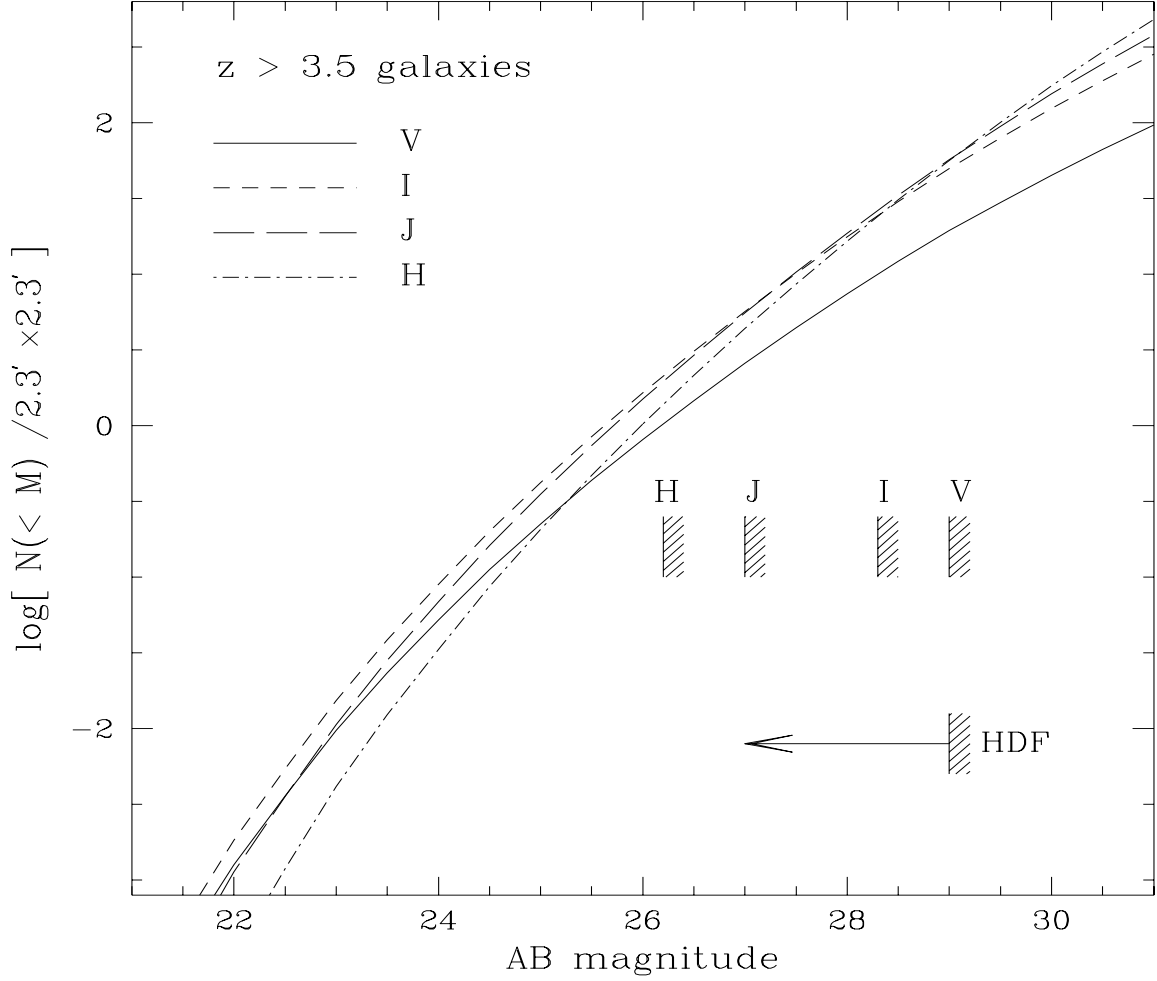}
\vspace*{4.5in}
\caption[Spectra] {\label{fig:svijh} Theoretical $V$, $I$, $J$, and
$H$ counts for $z>3.5$ galaxies in our standard $\Lambda$CDM model
with $Z_{\rm Ly\alpha}=10^{-3}{\rm Z_\odot}$.  Since most galaxies at
$V<30$ mag may be resolved, this model could still be consistent with
the HDF images.}
\end{figure}

Future observations at longer infrared wavelengths such as the $J$ and
$H$ infrared bands, would be better suited to the study of the
$z\gsim4.5$ universe as they allow bigger volumes to be probed. In
Figure~\ref{fig:qvijh}, we show the expected number counts of quasars
in the $V$, $I$, $J$, and $H$ bands for the model which is consistent
with the optical HDF data. An increasingly larger number of objects
are predicted to be detected in the $I$, $J$, and $H$ bands as these
are affected by intervening absorption only at extreme redshifts.
These number counts could be somewhat reduced since faint, compact WFC
sources lose signal--to--noise when observed by NICMOS, due to the
poorer angular resolution of the infrared images.
Figure~\ref{fig:svijh} shows the expected number of galaxies in the
infrared bands.  At 30 mag, the gain in observing galaxies redward of
the $I$-band is not as significant as for quasars, because the
intrinsic flux of unreddened galaxies, unlike those of quasars,
decreases as a function of increasing wavelength at $z\sim4$.  The
redshifts of either type of sources can be obtained using a
straightforward analogy to the $B$ dropout technique, applied to the
$I-J$ vs.  $J-H$ color--color diagram shown in Figure \ref{fig:ijjh}.
Future deep observations with NICMOS in these bands of a field
comparable in size to the HDF would be able to either reveal tens of
ultra--high redshift quasars, or else place tighter constraints on
quasar models than currently possible using $V$ and $I$ data.

\acknowledgements

We thank Steve Beckwith for discussions and Martin Haehnelt and Martin
Rees for useful comments. ZH thanks G. Djorgovski for suggesting to
predict number counts of high redshift sources in different bands.  PM
acknowledges support by NASA through grants NAG5-4236 and
AR-06337.10-94A from the Space Telescope Science Institute.  AL
acknowledges support from NASA ATP grant NAG5-3085, and the Harvard
Milton fund.

\section*{REFERENCES}
{
\StartRef

\Ref Babul, A., \& Rees, M. J. 1992, MNRAS, 255, 346

\Ref Bajtlik, S., Duncan, R. C., \& Ostriker, J. P. 1988, ApJ, 327, 570

\Ref Baugh, C. M., Cole, S., Frenk, C. S. \& Lacey, C. G. 1997, preprint astro-ph/9703111

\Ref Bruzual, G., \& Charlot, S. 1996, in preparation.  The models are available from the anonymous ftp site gemini.tuc.noao.edu.

\Ref Calzetti, D., Kinney, A. L., \& Storchi-Bergmann, T. 1994, ApJ, 429, 582

\Ref Carroll, S. M., Press, W. H., Turner, E. L. 1992, ARA\&A, 30, 499

\Ref Cole, S., Arag\'on-Salamanca, A, Frenk, C., Navarro, J., Zepf, S. 1994, MNRAS 271, 781

\Ref Conti, A., Kennefick, J. D., Martini, P., \& Osmer, P. S. 1998, in preparation

\Ref Efstathiou, G. 1992, MNRAS, 256, 43

\Ref Efstathiou, G., \& Rees, M. J. 1988, MNRAS, 230, 5p

\Ref Eisenstein, D. J. 1997, ApJ, submitted, preprint astro-ph/9709054

\Ref Eisenstein, D.J., \& Loeb, A. 1995, ApJ, 443, 11

\Ref Elston, R. A. W., Santiago, B. X., \& Gilmore, G. F. 1996, New Astronomy, 1, 1

\Ref Elvis, M., Wilkes, B. J., McDowell, J. C., Green, R. F., Bechtold, J., Willner, S. P., Oey, M. S., Polomski, E., \& Cutri, R. 1994, ApJS, 95, 1

\Ref Flynn, C., Gould, A., \& Bahcall, J. N. 1996, ApJ, 466, L55

\Ref Frank, J., King., A., \& Raine, D. 1992, Accretion Power in Astrophysics, Cambridge Univ. Press, Cambridge

\Ref Franx, M, Illingworth, G. D., Kelson, D. D., van Dokkum, P. G., \& Tran, K.-V. 1997, ApJ, 486, L75

\Ref Gnedin, N. Y., \& Ostriker, J. P. 1997, ApJ, 486, 581

\Ref Governato, F., Baugh, C. M., Frenk, C. S., Cole, S., Lacey, C. G., Quinn, T., \& Stadel, J. 1998, Nature, 392, 359

\Ref Haehnelt, M. G., \& Rees, M. J. 1993, MNRAS, 263, 168

\Ref Haehnelt, M. G., Natarajan, P., \& Rees, M. J. 1998, MNRAS, submitted, preprint astro-ph/9712259

\Ref Haiman, Z., \& Loeb, A. 1997, ApJ, 483, 21 (HL97)

\Ref Haiman, Z., \& Loeb, A. 1998a, in Science with the Next Generation Space Telescope, eds. E. Smith \& A. Koratkar, preprint astro-ph/9705144

\Ref Haiman, Z., \& Loeb, A. 1998b, ApJ, in press, astro-ph/9710208 (HL98)

\Ref Haiman, Z., Rees, M. J., \& Loeb, A. 1997, ApJ, 476, 458

\Ref Ikeuchi, S. 1986, Ap\&SS, 118, 509

\Ref Kauffmann, G., Guiderdoni, B., \& White, S. D. M. 1994, MNRAS, 267, 981

\Ref Kauffmann, G., \& White, S. D. M. 1993, MNRAS, 261, 921

\Ref Kormendy, J., Bender, R., Magorrian, J., Tremaine, S., Gebhardt, K., Richstone, D., Dressler, A., Faber, S. M., Grillmair, C., \& Lauer, T. R. 1997, ApJ, 482, L139

\Ref Lacey, C., \& Cole, S. 1993, MNRAS, 262, 627

\Ref Loeb, A. 1998, in Science with the Next Generation Space Telescope, eds. E. Smith \& A. Koratkar, preprint astro-ph/9704290

\Ref Madau, P. 1995, ApJ, 441, 18

\Ref Madau, P., Ferguson, H. C., Dickinson, M. E., Giavalisco, M., Steidel, C. C., Fruchter, A. 1996, MNRAS, 283, 1388

\Ref Magorrian, J., et al. 1997, submitted to AJ, preprint astro-ph/9708072

\Ref McLeod, K. 1996, in ESO-IAC conference on Quasar Hosts, Tenerife, in press

\Ref M\'{e}ndez, R. A., Minniti, D., De Marchi, G., Baker, A., \& Couch, W. J. 1996, MNRAS, 283, 666

\Ref Narayan, R. 1996, ``Advective Disks'', to appear in Proc. IAU Colloq. 163 on Accretion Phenomena \& Related Outflows, A.S.P. Conf. Series, eds. D. T. Wickramasinghe, L.  Ferrario, G. V. Bicknell, in press, preprint astro-ph/9611113

\Ref Navarro, J. F., Frenk, C. S., \& White, S. D. M. 1995, MNRAS, 275, 56

\Ref Navarro, J. F., \& Steinmetz, M. 1997, ApJ, 478, 13

\Ref Ostriker, J. P., \& Steinhardt, P. J. 1995, Nature, 377, 600

\Ref Pei, Y. C. 1995, ApJ, 438, 623

\Ref Pettini, M., Steidel, C. C., Adelberger, K. L., Kellogg, M., Dickinson, M. E., Giavalisco, M. 1997, to appear in `ORIGINS', ed. J.M. Shull, C.E. Woodward, and H. Thronson, (ASP Conference Series), preprint astro-ph/9708117

\Ref Press, W. H., \& Schechter, P. L. 1974, ApJ, 181, 425

\Ref Quinn, T., Katz, N., \& Efstathiou, G. 1996, MNRAS, 278, 49

\Ref Sasaki, S. 1994, PASJ, 46, 427

\Ref Scalo, J. M. 1986, Fundamentals of Cosmic Physics, vol. 11, p. 1-278

\Ref Schaller, G., Schaerer, D., Meynet, G., \& Maeder, A. 1992, A\&ASS, 96, 269

\Ref Schneider, D. P., Schmidt, M., \& Gunn, J. E. 1991, AJ, 102, 837

\Ref Small, T. A., \& Blandford, R. D. 1992, MNRAS, 259, 725

\Ref Smith, E. P., \& Koratkar, A. eds, 1998, Science with the Next Generation Space Telescope, Ast. Soc. of the Pacific Conf. Series, v. 133

\Ref Somerville, R. S., Primack, J. R. 1998, MNRAS, submitted 

\Ref Songaila, A. 1997, ApJ, 490, L1

\Ref Songaila, A., \& Cowie, L. L. 1996, AJ, 112, 335

\Ref Steidel, C. C., Adelberger, K. L., Dickinson, M. E., Giavalisco, M, Pettini, M., Kellogg, M. 1998, ApJ, 492, 428

\Ref Thoul, A.A., \& Weinberg, D.H. 1996, ApJ, 465, 608

\Ref Tytler, D. et al. 1995, in QSO Absorption Lines, ESO Astrophysics Symposia, ed. G. Meylan (Heidelberg: Springer), p.289

\Ref van der Marel, R. de Zeeuw, P. T., Rix, H-W., \& Quinlan, G. D. 1997, Nature, 385, 610 
\Ref Viana, P. T. P., \& Liddle, A. R. 1996, MNRAS 281, 323

\Ref Vogeley, M. S. 1997, ApJ, in press, preprint astro-ph/9711209

\Ref Williams, R. E., et al. 1996, AJ, 112, 1335

\Ref Yi, I. 1996, ApJ, 473, 645

\Ref Zheng, W., Kriss, G. A., Telfer, R. C., Grimes, J. P., Davidsen, A. F. 1997,  ApJ, 475, 469

}


\clearpage
\newpage

\clearpage
\newpage

\clearpage
\newpage

\clearpage
\newpage

\clearpage
\newpage

\clearpage
\newpage

\end{document}